\documentstyle{aipproc}

\begin{document}
\title{Isospin Splitting in the Pion-Nucleon Couplings from QCD Sum Rules}

\author{Thomas Meissner$^*$ and Ernest Henley$^{\dagger}$}
\address{$^*$Department of Physics, Carnegie Mellon University,
Pittsburgh, PA 15213\thanks{email: meissner@yukawa.phys.cmu.edu}\\
$^{\dagger}$Department of Physics, University of Washington, Box 351560,
Seattle, WA 98195\thanks{email: henley@alpher.npl.washington.edu}}

\maketitle

\begin{abstract}
We use QCD sum rules for the three point function 
of a pseudoscalar and two nucleonic currents in order
to estimate the charge dependence of the pion nucleon coupling
constant $g_{NN\pi}$ coming from isospin violation in the strong
interaction. The effect can be attributed primarily to the difference 
of the quark condensates $<{\bar u}u>$ and $<{\bar d}d>$.
Assuming that the $\pi^0$ is a pure isostate we obtain for the splitting
$(|g_{pp\pi_0}| - |g_{nn\pi_0}|) / g_{NN\pi}$ 
an interval of $0.8 * 10^{-2}$ to $2.3 * 10^{-2}$,
the uncertainties coming mainly from the  input parameters.
In order to obtain the coupling to a physical $\pi^0$ we have 
to take $\pi - \eta$ mixing into account leading to an interval of 
$1.2 * 10^{-2}$ to $3.7 * 10^{-2}$.
The charged pion nucleon coupling is found to be the average of
$| g_{pp\pi_0}|$ and $|g_{nn\pi_0}|$. Electromagnetic effects are not included.
\end{abstract}

The effect of isospin violating meson nucleon couplings 
is of great importance 
in the investigation of charge
symmetry breaking (CSB) phenomena \cite{MNS}.
On a microscopical level, isospin symmetry is broken by the electromagnetic
interaction as well as the mass difference of up and down quarks
$m_u \ne m_d$.
We will examine
the splitting between the
pion nucleon coupling constants 
$g_{pp\pi^0}$, $g_{nn\pi^0}$ and $g_{pn\pi^+}$ using the QCD sum rule 
method, 
which has been established as a powerful and fruitful technique for describing hadronic 
phenomena at intermediate energies.
We will only look at effects which arise from isospin breaking 
in the strong interaction. 
In the QCD sum rule method this is reflected by
$m_u \ne m_d$ as well as by the isospin breaking of the vacuum condensates.
Electromagnetic effects are  not examined.
Details of our calculation have been recently published \cite{MH}.

We start from the three point function of two nucleonic (Ioffe) currents ($\eta_N$)
and one pseudoscalar isovector ($P^{T=1} _a$) interpolating current with the appropriate isospin
quantum numbers:
\begin{equation}
A_{NN\pi^a} (p_1,p_2,q) = 
\int d^4 x_1 d^4 x_2 e^{ip_1 x_1} e^{- ip_2 x_2}
\left \langle 0 \vert {\cal T} 
\eta_N (x_1) P^{T=1} _a (0) {\bar{\eta_N}} (x_2) \vert 0 \right \rangle ,
\label{eq1}
\end{equation}
where $a$ stands for $+$ or $0$ and $N$ for proton or neutron, respectively.
The momenta $p_1$ and $p_2$ are those of the nucleon, and 
$q = p_1 - p_2$ that of the pion. 
We are only keeping terms up to first order
in isospin violation, i.e. $m_d - m_u$.

The phenomenological side of the QCD sum rules 
for the three point functions $A$ are obtained by saturating the 
general expressions for the $A$'s (\ref{eq1}) with the 
corresponding nucleon and pion intermediate states.
The overlap between the pion states and the pseudoscalar interpolating fields
are given by  current algebra and the axial Ward identity.
For $a=0$ and $N=p$ we have for example:
\begin{equation}
A_{pp\pi^0} = \, i \, {\lambda_p}^2 \, 
\frac{{m_\pi}^2 f_\pi}{m_0} \,
\frac{ g_{pp\pi^0}}{-q^2 + {m_\pi}^2} \,
\frac{1}{{p_1}^2 - {M_p}^2} \,
\frac{1}{{p_2}^2 - {M_p}^2} \,
M_p \, \gamma_5 \, \not{\! q} \,
\,\, + \,\, \dots ,  
\label{eq6} 
\end{equation}
where the $\lambda_p$ denotes the overlap between the proton Ioffe 
current and the corresponding single nucleon state.
The difference between $\lambda_p$ and $\lambda_n$ is an input parameter
for the sum rule analysis.
It can be eliminated by using the isospin violating sum rule 
for the nucleon 2 point function which makes a prediction for the proton-neutron
mass splitting \cite{YHHK}.
The $\dots$ denote contributions from higher resonances
and the continuum.

The three point function method works at large spacelike
$q^2 \approx 1\mbox{GeV}^2$.
We neglect $m_\pi^2$ and identify the 
pion pole term $\frac{1}{q^2}$ on the r.h.s. of 
eq.(\ref{eq6}) with the corresponding $\frac{1}{q^2}$
term in the operator product expansion (OPE) \cite{RRY1}.
This treatment is justified in the isospin conserving
case, where it has been shown that the non-$\frac{1}{q^2}$
terms give rise to a pion nucleon form factor
$g_{\pi NN} (q^2)$ with a monopole cutoff mass of about
$800 \mbox{MeV}$  \cite{Mei}, which is almost identical to the result of 
lattice calculations.

In the OPE we include only the leading order term, which in this case 
is the the quark condensate $<\bar{q} q>$.
The isospin splitting of the higher order condensates is practically unknown,
the error due to their omission can be estimated to about $25\%$ of the leading order
contribution. 

The Borel sum rule for the $\not{\! q} \gamma_5$ structure, which has a double nucleon pole,
gives an interval of
\begin{equation} 
8 * 10^{-3} < \left ( \frac{(|g_{pp\pi^0}| - |g_{nn\pi^0} |)}{g_{NN\pi}} \right ) < 23 * 10^{-3} .
\label{interval1}
\end{equation}
where $g_{NN\pi} \equiv \frac{1}{2} ( |g_{pp\pi^0}| + |g_{nn\pi^0} |)$. 

A crucial input parameter is the isospin violation 
in the quark condensate $\gamma = \frac{<{\bar d}d>}{<{\bar u}u>} - 1$. 
The range for $\gamma$ obtained from various other methods is rather large: 
$0.002 < - \gamma < 0.010$. 
This leads to a significant uncertainty in the final
result for $(|g_{pp\pi^0}| - |g_{nn\pi^0} |)$.

Up to now we have treated the $\pi^0$ as pure isospin state.
If we want to calculate the coupling to the physical pion we have to include
$\pi-\eta$ mixing, which can be done at tree level chiral perturbation theory.
The final result is:
\begin{equation} 
12 * 10^{-3} < \left ( \frac{(|g_{pp\pi^0}| - |g_{nn\pi^0} |)}{g_{NN\pi}} \right ) < 37 * 10^{-3} .
\label{interval2}
\end{equation}
It should be noted that in charge symmetry breaking NN potential calculations
it is common to treat the $\pi-\eta$ mixing separately and to 
use the coupling to the pure isostate $\pi^0$.

Up to first order in isospin breaking
the charged pion nucleon coupling is exactly the arithmetic average
of the two neutral pion nucleon couplings:
\begin{equation}
\frac{|g_{pn\pi^+}|}{\sqrt{2}} 
= \frac{1}{2} [ |g_{pp\pi^0}| + |g_{nn\pi^0}| ] .
\label{eq9}
\end{equation}
which is a simple consequence of the $u$ and $d$ quark contents of
the three point functions and valid within the approximations considered.

Our result can be compared with those of other approaches \cite{MNS}
which 
analyze the isospin splitting of the pion nucleon couplings arising from 
the strong interaction, i.e.,  essentially
the quark mass difference $m_d -m_u$ (Table \ref{tab1}).

\begin{table}[ht]
\caption{ }
\begin{tabular}{cc}  &
$({|g_{pp{\pi_0}}| - |g_{nn{\pi_0}}}|) / {g_{NN\pi}}$  \\ \tableline
this work & $\approx 0.012 \dots 0.037$ \\
Quark-Gluon Model  &  $\approx 0.010 \dots 0.014$ \\ 
Quark-Pion Model   & $\approx 0.006$  \\
$\pi-\eta$ mixing & $ 0.005 \pm 0.0018$ \\ 
Chiral Bag Model  & $ \approx 0.0067$ \\
Cloudy Bag Model  & $\approx - 0.006$ \\
\end{tabular}
\label{tab1}
\end{table}
Direct experimental values are not available.
The Nijmegen phase shift analysis for $NN$ and 
$N {\bar N}$ scattering data,
which includes electromagnetic effects, 
finds $\left ( \frac{(|g_{pp\pi^0}| - |g_{nn\pi^0} |)}
{g_{NN\pi}} \right )$ = 0.002, but with an 
error of 0.008; thus, there is no evidence for a difference and it also 
find no evidence for a difference between $g_{pn\pi^+}$ and $g_{NN\pi^0}$
within the statistical errors \cite{Nmw}.

\end{document}